\documentclass[12pt,chicago,iop]{emulateapj}

\usepackage{amsmath,natbib,graphicx}
\usepackage{epsf}
\usepackage{fancyvrb}
\usepackage{epstopdf}
\usepackage{epsfig}
\usepackage{color}
\DeclareGraphicsExtensions{.jpg,.pdf,.png,.eps,.ps}

\begin{document}
\VerbatimFootnotes

\title{A proposal for climate stability on H$_2$-greenhouse planets}
\shorttitle{H$_2$-greenhouse climate stability}


\author{Dorian~S.~Abbot\altaffilmark{1}}
\altaffiltext{1}{Department of the Geophysical Sciences, University of
  Chicago, 5734 South Ellis Avenue, Chicago, IL 60637}
\shortauthors{Abbot}

\email{abbot@uchicago.edu}


\begin{abstract}
  A terrestrial planet in an orbit far outside of the standard
  habitable zone could maintain surface liquid water as a result of
  H$_2$-H$_2$ collision-induced absorption by a thick H$_2$
  atmosphere. Without a stabilizing climate feedback, however,
  habitability would be accidental and likely brief. In this letter I
  propose stabilizing climate feedbacks for such a planet that require
  only that biological functions have an optimal temperature and
  operate less efficiently at other temperatures. For example, on a
  planet with a net source of H$_2$ from its interior, H$_2$-consuming
  life (such as methanogens) could establish a stable climate. If a
  positive perturbation is added to the equilibrium temperature, H$_2$
  consumption by life will increase (cooling the planet) until the
  equilibrium climate is reestablished. The potential existence of
  such feedbacks makes H$_2$-warmed planets more attractive
  astrobiological targets.
\end{abstract}

\bigskip\bigskip

\section{Introduction}

The traditional definition of planetary habitability is the ability of
a planet to maintain liquid water at its surface. For a planet to be
of astrobiological interest, it must maintain habitability for
timescales relevant for biological macroevolution (at least tens of
millions of years). Earth has been able to maintain habitable
conditions for most of the last four billion years despite the solar
flux increasing by about 50\% over that period
\citep{SAGAN:1972p1233}. A possible explanation for this is the
silicate-weathering feedback, which would regulate the atmospheric
CO$_2$ in such a way as to maintain habitable conditions as other
forcings, such as solar flux, varied
\citep{Walker-Hays-Kasting-1981:negative}. The silicate-weathering
feedback is also essential for the standard habitable zone. The
standard habitable zone is defined as the circumstellar region where
an Earth-like planet (having both land and ocean, volcanism, and a
predominantly N$_2$-CO$_2$-H$_2$O atmosphere) could maintain habitable
conditions, assuming the functioning of the silicate-weathering
feedback \citep{Kasting93}.

Recent work has pointed out that planets outside of the standard
habitable zone with thick H$_2$ atmospheres could maintain surface
liquid water due to the greenhouse effect of H$_2$-H$_2$
collision-induced absorption
\citep{Stevenson99,Pierrehumbert:2011p3366}. This is noteworthy
because many terrestrial planets larger than Earth (super-Earths) have
been detected, and calculations suggest that they could retain an
H$_2$ atmosphere (against escape to space) for billions of years.
Potential biosignatures on this type of planet have already been
investigated \citep{Seager:2013fi}. The habitability of such a planet,
however, would be transient or accidental in the absence of some
stabilizing feedback \citep{Wordsworth2012-transient}, rendering such
planets significantly less interesting places to search for life.

\citet{Wordsworth2012-transient} notes that the reducing atmosphere of
an H$_2$-greenhouse planet would produce pre-biotic compounds that
would favor the origin of life and speculates that this life could
evolve mechanisms to regulate the H$_2$ pressure and maintain
habitable surface conditions. This is in contrast with
\citet{Pierrehumbert:2011p3366}, who write that the evolution of life
on an H$_2$-greenhouse planet would likely lead to the consumption
of H$_2$ and the destruction of the atmosphere (and habitability). A
factor that could affect this conclusion of
\citet{Pierrehumbert:2011p3366} is that it seems likely that the
destruction of H$_2$, which supports the conditions that allow life,
by life itself would inherently be self-limiting. The arguments in this
letter will be built around this idea.

The purpose of this letter is to propose specific biological feedbacks
that could maintain habitable conditions on a planet with an
H$_2$-dominated atmosphere and greenhouse warming provided primarily
by H$_2$-H$_2$ collision-induced absorption. Such feedbacks would
allow the habitable zone for this type of planet to extend to very
large distances from its host star, greatly expanding the list of
attractive candidates for biosignature searches. This work is
speculative, yet I think worthwhile because I approach it by making
a minimum of reasonable assumptions and following them to their
logical conclusion. I will begin by outlining the critical assumptions
made in this letter (section~\ref{sec:assumptions}). Next I will show
how a stabilizing feedback would work on planets with a net source of
H$_2$ from the interior to the atmosphere (section~\ref{sec:source})
and on planets with a net loss of H$_2$ to space
(section~\ref{sec:sink}). Finally, I will discuss these results
(section~\ref{sec:discussion}) and conclude
(section~\ref{sec:conclusions}).

\section{Important Assumptions}
\label{sec:assumptions}

\begin{figure}[h]
\begin{center}
\epsfig{file=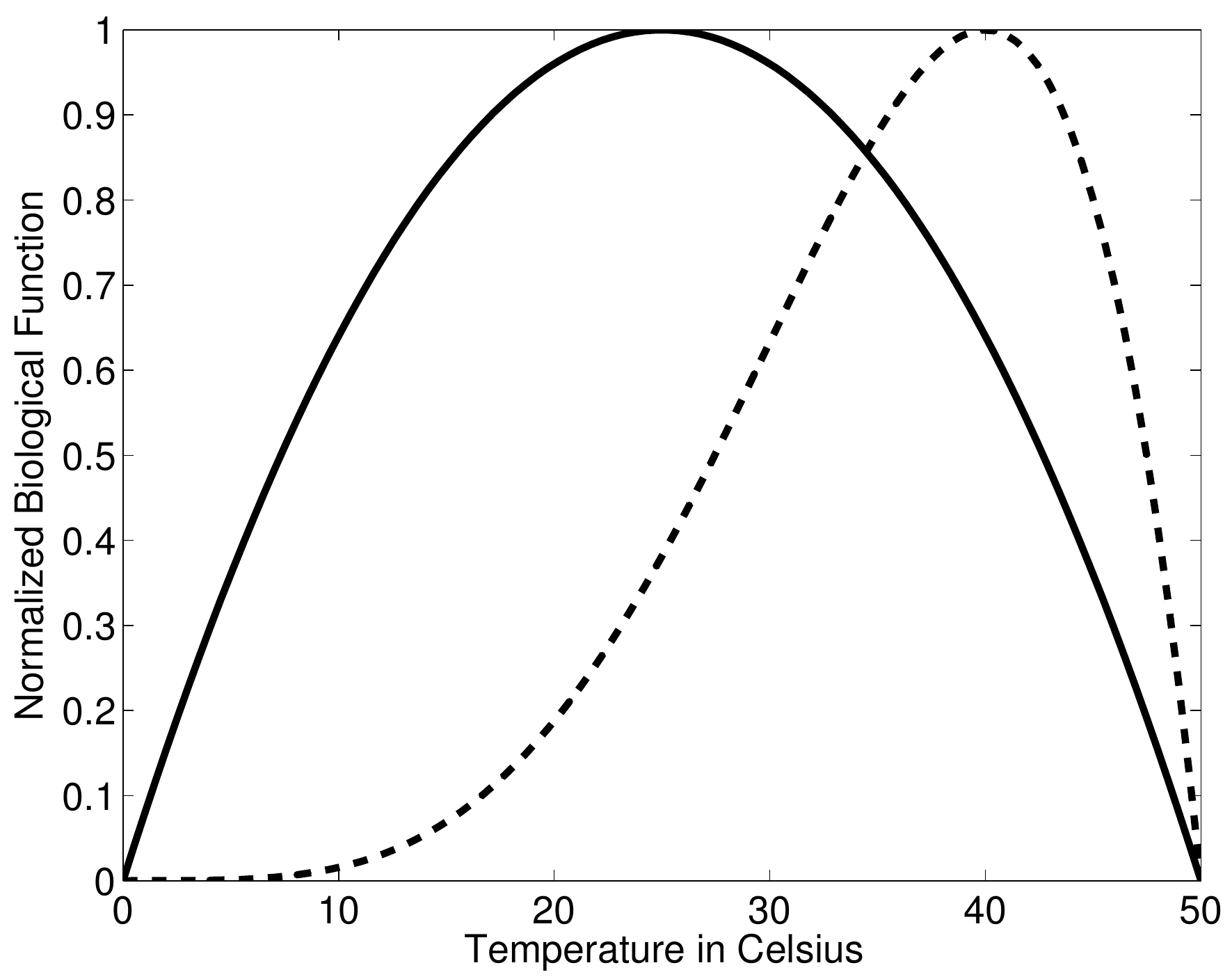, width=8cm} 
\end{center}
\caption{These are example curves of normalized biological functioning
  (e.g., production or destruction of H$_2$) as a function of
  planetary surface temperature. The solid curve shows the relation
  plotted in Equation~(\ref{eq:temp-dep}), which is symmetric around
  the temperature of maximum biological functioning. The dotted curve
  represents an alternative temperature dependence of biological
  functioning in which the decrease in efficiency at high temperatures
  is steeper than the increase in efficiency at low temperatures. In
  this letter I use the symmetric curve because it simplifies the
  mathematics, but the asymmetric curve is more realistic in many
  cases and will yield qualitatively similar results.}
\label{fig:function}
\end{figure}

The most important assumption I will make is that biological processes
are temperature-dependent such that there is a maximum rate of
functioning at some temperature (and consequently functioning
decreases at temperatures above or below this temperature,
Figure~\ref{fig:function}). This assumption describes the behavior of
many terrestrial biological systems, and it is difficult to think of a
realistic biological system that would not obey it. An illustrative
functional form that describes this type of temperature dependence is
\begin{equation}
  F(T)=F_0\frac{(T-T_1)[(T_2-T_1)-(T-T_1)]}{(T_2-T_1)^2},
  \label{eq:temp-dep}
\end{equation}
where $F(T)$ is the rate of functioning (which will correspond to
production or destruction of H$_2$ in this letter), $F_0$ is a base
functioning rate dependent on factors such as nutrient availability,
$T_1$ is the minimum temperature for the biological function, and
$T_2$ is the maximum temperature for the biological function. I will
assume that the biological function does not operate outside of the
temperature range $T_1 \leq T \leq T_2$.

I will also assume that we are considering planets distant enough from
their host star that they have a very low emission temperature (60K or
lower). In such cases H$_2$ is optically thick near the radiating
temperature, and other greenhouse gases such as CO$_2$, CH$_4$, and
H$_2$O are effectively not present at that atmospheric pressure
because their vapor pressure is too low
\citep{Pierrehumbert:2011p3366}. These gases therefore cannot affect
the outgoing longwave radiation. It is a reasonable approximation to
take the adiabatic lapse rate as independent of these minor
constituents, so that H$_2$ is the only greenhouse gas relevant for
determining the surface temperature.

\section{Planets with a net hydrogen source from the interior}
\label{sec:source}

I will first consider planets with a net source of hydrogen from their
interior, either from direct outgassing or as a result of
serpentinization. One example would be an Earth-like planet with a
large hydrogen envelope in interstellar space \citep{Stevenson99}. If
such a planet has the right amount of H$_2$ for liquid water on its
surface, then life can evolve on it. One likely option is that
methanogenic life (henceforth methanogens) will take advantage of
available H$_2$ through a reaction such as
\begin{equation}
  \text{CO$_2$+4H$_2$} \longrightarrow \text{CH$_4$+2H$_2$O}, 
  \label{eq:methanogen}
\end{equation}
consuming H$_2$ in the process. The CO$_2$ in
Equation~(\ref{eq:methanogen}) could be outgassed from the interior if
it is relatively oxidized, or it could be left over from formation. It
is also possible that life could reduce some element other than carbon
to obtain energy. What is important is that H$_2$ is consumed during
the process. If we assume that these methanogens function with a
temperature dependence given by Equation~(\ref{eq:temp-dep}), then we
can write the following equation for the time derivative of the
hydrogen pressure
\begin{equation}
  \frac{d p_{H_2}}{d t} = P_0-4L_0\frac{(T-T_1)[(T_2-T_1)-(T-T_1)]}{(T_2-T_1)^2},
  \label{eq:pressure1}
\end{equation}
where $P_0$ is the background source rate of H$_2$ from the interior,
and $L_0$ is the scale of H$_2$ consumption by methanogens, which is
dependent on factors such as nutrient availability. I have inserted
the factor of four in Equation~(\ref{eq:pressure1}) for later
mathematical convenience.

The assumption that H$_2$ is the only greenhouse gas that affects
outgoing longwave radiation \citep{Pierrehumbert:2011p3366} leads to
an equation yielding the mean surface temperature as a function of the
H$_2$ pressure. Following the fit to radiative calculations given in
\citet{Wordsworth2012-transient}, I will use the following equation
\begin{equation}
  \frac{1}{4}(1-\alpha)S_0+F_{geo}= a p_{H_2}^{-1} T^4,
\label{eq:climate}
\end{equation}
where $T$ is the planetary surface temperature, $\alpha$ is the
planetary albedo, $S_0$ is the stellar constant, $F_{geo}$ is the
geothermal heat flux, and $a$ is a constant that is defined slightly
differently here than in \citet{Wordsworth2012-transient}. For an
Earth-gravity planet, $a=7.3 \times 10^{-9}$~W~m$^{-2}$~K$^{-4}$~Pa. The
exponent of $p_{H_2}$ in Equation~(\ref{eq:climate}) is actually
slightly different from negative one. I approximate it as negative one
because doing so simplifies the mathematics and does not change the
qualitative behavior of the system. I assume that the planetary albedo
is constant, which also does not change the qualitative behavior of
the system since increasing $p_{H_2}$ leads to significantly more
greenhouse warming than it increases Rayleigh scattering
\citep{Wordsworth2012-transient}, particularly for planets in distant
orbits.

Let us define $F=\frac{1}{4}(1-\alpha)S_0+F_{geo}$ and assume that $F$
is constant on the timescale of biological adjustment of H$_2$ (we
will revisit this assumption in section~\ref{sec:discussion}). We can
then linearize Equation~\ref{eq:climate} around $T_1$ and take the
time derivative to find that
\begin{equation}
  4 a T_1^3 \frac{dT}{dt} = F \frac{d p_{H_2}}{d t},
\label{eq:dptodT}
\end{equation}
when $T \approx T_1$. By combining
Equations~(\ref{eq:pressure1})~and~(\ref{eq:dptodT}), defining a
nondimensional temperature $\phi=\frac{T-T_1}{T_2-T_1}$, defining a
nondimensional time $\tau=t\frac{L_0 F}{4 a T_1^3(T_2-T_1)}$, and
defining a nondimensional hydrogen outgassing rate
$\beta = \frac{P_0}{L_0}$, we find the following
nondimensional equation for surface temperature of the planet
\begin{equation}
\frac{d\phi}{d\tau} = \beta - 4 \phi(1-\phi).
\label{eq:nondim1}
\end{equation}

\begin{figure}[h]
\begin{center}
\epsfig{file=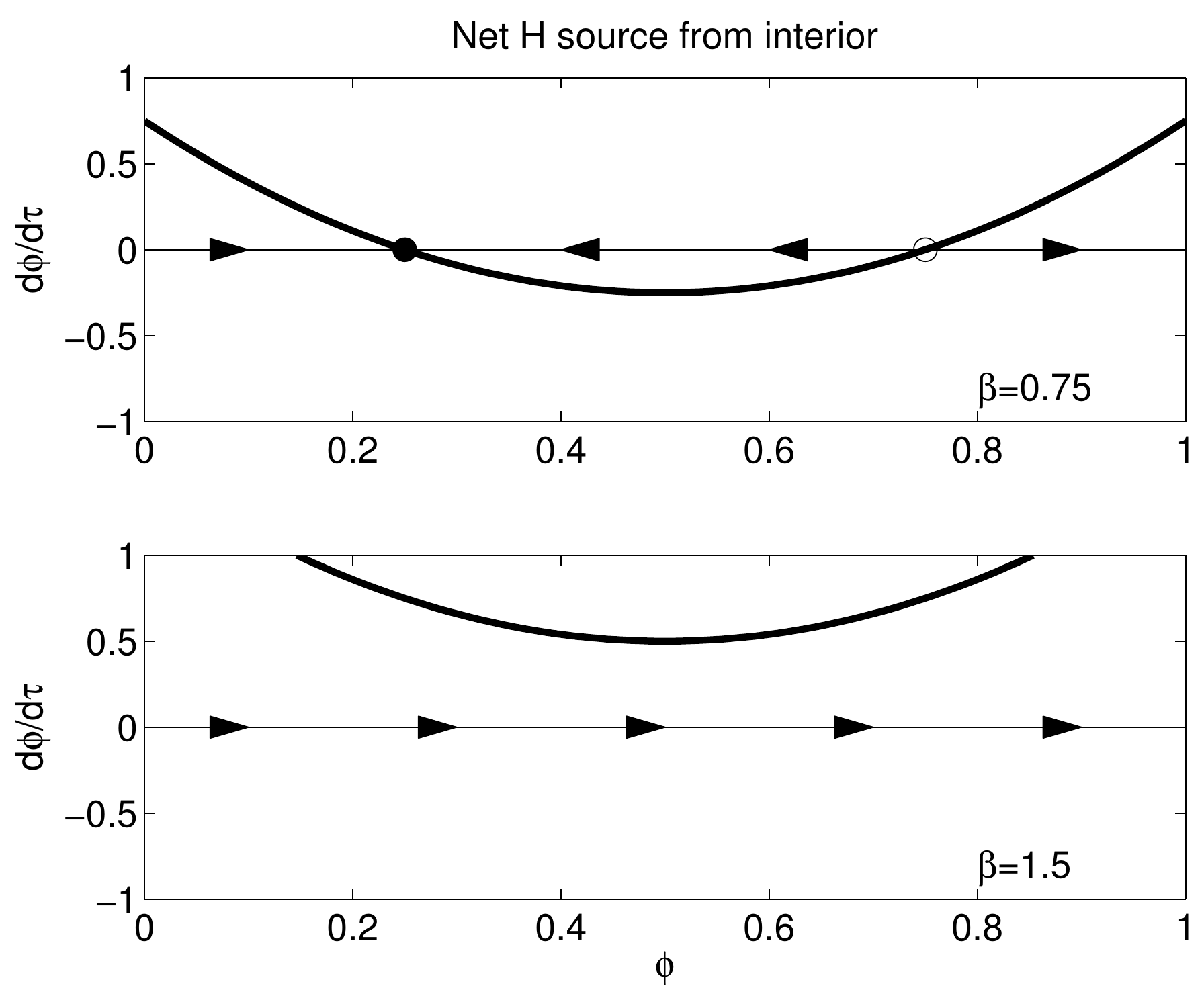, width=8cm} 
\end{center}
\caption{Diagrams showing the nondimensional rate of change of
  temperature ($\frac{d\phi}{d\tau}$) of an H$_2$-greenhouse planet
  with a net source of hydrogen from its interior as a function of the
  nondimensional temperature ($\phi$) for two different values of the
  nondimensional hydrogen outgassing rate ($\beta$). Arrows along the
  $\frac{d\phi}{d\tau}=0$ line indicate the flow direction of $\phi$,
  as determined by the sign of $\frac{d\phi}{d\tau}$. Circles where
  $\frac{d\phi}{d\tau}$ intersects zero for $\beta<1$ (in the top
  panel) are fixed points, and represent temperatures where the
  climate achieves an equilibrium. The open circle is an unstable
  fixed point, and the filled circle is a stable fixed point. The
  climate returns to the stable fixed point when it is perturbed away
  from it. If $\beta$ is increased above one (bottom panel), a
  bifurcation occurs and the system no longer has any fixed points. If
  the asymmetric biological temperature dependence curve from
  Figure~\ref{fig:function} were used, the qualitative behavior would
  be the same, but the stable and unstable fixed points would not be
  symmetric around $\phi=0.5$.}
\label{fig:hab_ode_plot1}
\end{figure}

When $\beta<1$, Equation~(\ref{eq:nondim1}) has two steady-state
solutions (fixed points, Figure~\ref{fig:hab_ode_plot1}). The solution
with $\phi<0.5$ is stable and the solution with $\phi>0.5$ is
unstable. The stable solution is associated with the increase in
methanogenic H$_2$ consumption with temperature at low temperatures
(Figure~\ref{fig:function}). If some perturbation increases the
planetary temperature from its equilibrium value, methanogens consume
H$_2$ more efficiently (reducing the greenhouse effect and cooling the
planet) and decrease the temperature back to its stable steady-state
value. By this mechanism an H$_2$-greenhouse planet with a net source
of hydrogen from its interior can maintain climate stability against
changes in the stellar flux or geothermal heat flux on geological
timescales if methanogens have evolved on it. If $\beta$ is increased
above one, a bifurcation occurs and the system no longer has any fixed
points. Instead, the temperature increases indefinitely. This
corresponds to a hydrogen outgassing rate so high that methanogens
cannot consume H$_2$ fast enough to allow a stable steady-state
climate. Such a situation could perhaps be due to nutrient limitation
of the methanogen population.

\section{Planets with net hydrogen loss to space}
\label{sec:sink}

Let us now consider a planet with a net loss of hydrogen to space. One
example might be a distant H$_2$-greenhouse planet orbiting an active
young star. Let us assume that as the surface conditions evolve on our
hypothetical planet as a result of H$_2$ loss to space, habitable
conditions are reached, and life evolves
\citep{Wordsworth2012-transient}. Some H$_2$-greenhouse planets could
have enough photosynthetically active radiation reach their surface to
sustain photosynthesis \citep{Pierrehumbert:2011p3366}. Let us assume
that a biosphere evolves on the planet that is capable of producing
biomass through photosynthesis that releases H$_2$ as a byproduct. The
following reaction is one example of this type of photosynthesis
\begin{equation}
  \text{CH$_4$+H$_2$O+h$\nu$}\longrightarrow\text{``CH$_2$O''+2H$_2$}, 
  \label{eq:biomass}
\end{equation}
where ``CH$_2$O'' likely represents more complex organic molecules on
which carbon has a net oxidation state of zero. On this reduced planet
photosynthesis would use solar energy to oxidize carbon and produce
biomass. It is also possible that methanogens would evolve that would
consume H$_2$ through Equation~(\ref{eq:methanogen}) or by running
Equation~(\ref{eq:biomass}) backwards (a reducing kind of respiration)
to obtain energy by reducing organic carbon, leading to cycling
between H$_2$ production and destruction. We can think of
Equation~(\ref{eq:biomass}) as either the only relevant reaction, or
we can imagine that it is the net effect of photosynthesis,
methanogenic H$_2$ consumption, and either the burial or other
disposal of organic matter. Since we are considering a habitable
planet, it is reasonable to assume that there is abundant H$_2$O on
the surface. I will assume that CH$_4$ is obtained from the planetary
interior through outgassing, or potentially through serpentinization
reactions if CO$_2$ is preferentially outgassed \citep{Oze:2005ch}.
The latter process uses H$_2$ to create CH$_4$, so for the feedback I
will describe to function it would have to create an overabundance of
CH$_4$ which Equation~(\ref{eq:biomass}) could then convert back to
H$_2$ as needed.

Following similar logic to that in section~\ref{sec:source}, including
assuming that biological production of H$_2$ by
Equation~(\ref{eq:biomass}) has a temperature dependence given by
Equation~(\ref{eq:temp-dep}), we can derive the following equation for
the time derivative of the hydrogen pressure on such a planet
\begin{equation}
  \frac{d p_{H_2}}{d t} = 4P_0\frac{(T-T_1)[(T_2-T_1)-(T-T_1)]}{(T_2-T_1)^2} - L_0,
  \label{eq:pressure}
\end{equation}
where $L_0$ is the net loss of H$_2$ to space unrelated to biological
H$_2$ production and $P_0$ is the scale of biological H$_2$
production. $L_0$ is increased by thermal and hydrodynamic escape to
space, large meteorite impacts that knock some atmosphere off the
planet, and solar wind. $L_0$ is decreased by fluxes of H$_2$ from the
interior to the atmosphere as a result of direct outgassing and
serpentinization. I will assume that $L_0$ does not depend on surface
temperature.

We can combine Equation~(\ref{eq:pressure}) with
Equation~(\ref{eq:dptodT}) and nondimensionalize to find
\begin{equation}
  \frac{d \phi}{d \tau} = 4 \phi (1-\phi)-\gamma,
  \label{eq:nondim2}
\end{equation}
where $\phi=\frac{T-T_1}{T_2-T_1}$ is again the nondimensional
temperature, $\tau=t\frac{P_0 F}{4 a T_1^3(T_2-T_1)}$ is the
nondimensional time, and $\gamma = \frac{L_0}{P_0}$ in the
nondimensional loss rate of hydrogen to space.

\begin{figure}[h]
\begin{center}
\epsfig{file=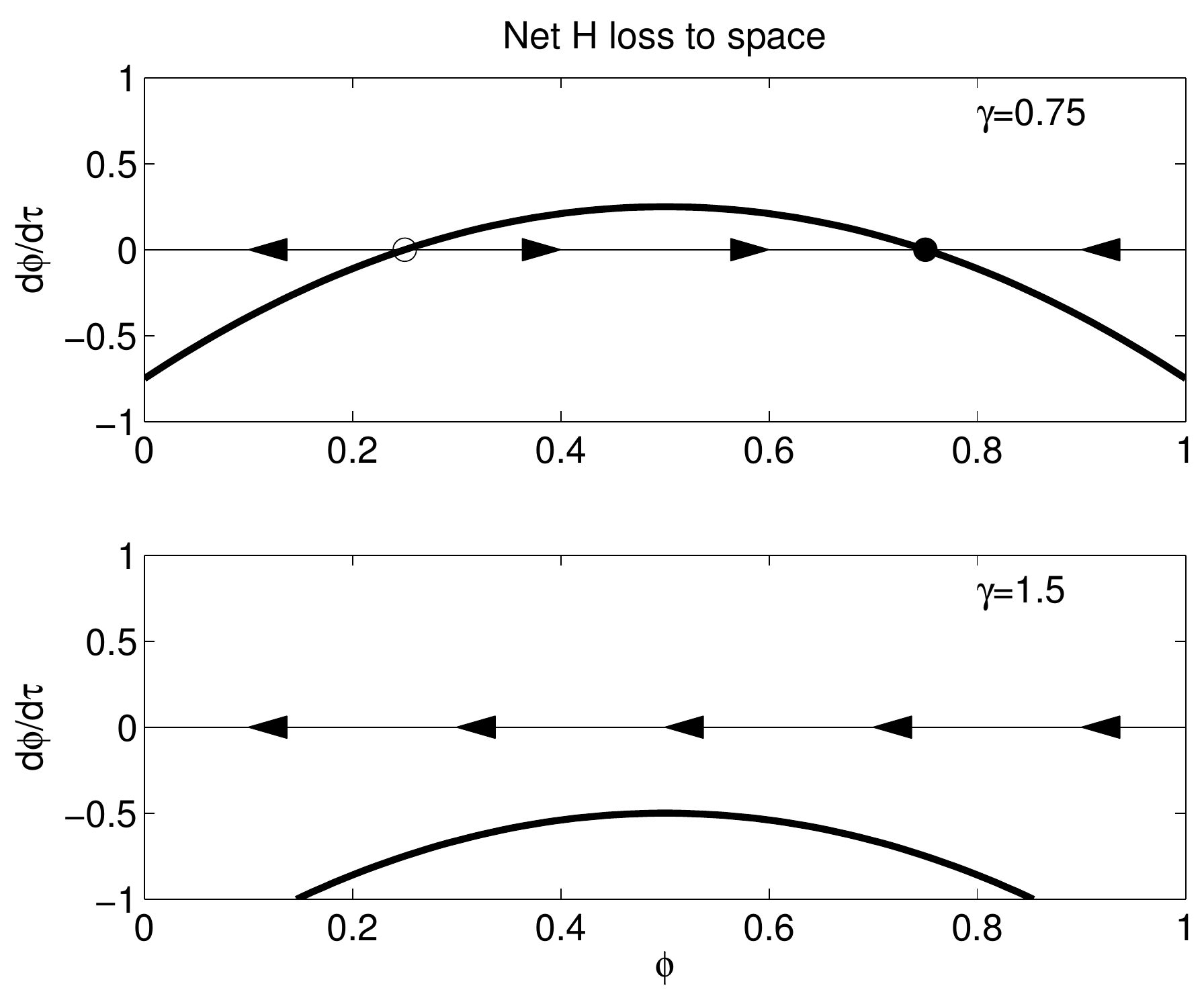, width=8cm} 
\end{center}
\caption{Diagrams showing the nondimensional rate of change of
  temperature ($\frac{d\phi}{d\tau}$) of an H$_2$-greenhouse planet
  with net loss of hydrogen to space as a function of the
  nondimensional temperature ($\phi$) for two different values of the
  nondimensional loss rate of hydrogen to space ($\gamma$). Arrows
  along the $\frac{d\phi}{d\tau}=0$ line indicate the flow direction
  of $\phi$, as determined by the sign of $\frac{d\phi}{d\tau}$.
  Circles where $\frac{d\phi}{d\tau}$ intersects zero for $\beta<1$
  (in the top panel) are fixed points, and represent temperatures
  where the climate achieves an equilibrium. The open circle is an
  unstable fixed point, and the filled circle is a stable fixed point.
  The climate returns to the stable fixed point when it is perturbed
  away from it. If $\gamma$ is increased above one (bottom panel), a
  bifurcation occurs and the system no longer has any fixed points. If
  the asymmetric biological temperature dependence curve from
  Figure~\ref{fig:function} were used, the qualitative behavior would
  be the same, but the stable and unstable fixed points would not be
  symmetric around $\phi=0.5$.}
\label{fig:hab_ode_plot2}
\end{figure}

For $\gamma<1$ the system has two fixed points
(Figure~\ref{fig:hab_ode_plot2}). The fixed point with $\phi>0.5$ is
stable, which means that if the planet occupies this fixed point and
is perturbed away from it, it will tend to return to its original
temperature. This stable fixed point is associated with the loss of
efficiency of biological H$_2$ production at higher temperatures
(Equation~\ref{eq:temp-dep}). If the planet occupies this fixed point
and some perturbation leads to an increase in surface temperature, the
biological production of H$_2$ will decrease, which will decrease the
H$_2$ pressure (Equation~(\ref{eq:pressure})) and therefore decrease
the surface temperature back toward its starting point
(Equation~\ref{eq:climate}). As long as the net loss of hydrogen to
space is low enough ($\gamma<1$), the planet will be stable against
secular perturbations (e.g., increases in stellar flux) and random
perturbations (e.g., random changes in outgassing that lead to changes
in $L_0$) as long as they are small enough that the system is not
perturbed beyond its unstable fixed point. If, however, the net loss
of hydrogen to space becomes too large ($\gamma>1$), the system has no
fixed points (Figure~\ref{fig:hab_ode_plot2}) and the atmosphere
simply collapses as a result of H$_2$ escape to space, leading to
permanent loss of habitability. This could result from either nutrient
limitation of biological capacity, or an excessively high base rate of
H$_2$ loss to space.

\section{Discussion}
\label{sec:discussion}

The model I have derived here assumes that biological functions can
adjust H$_2$ faster than the 10$^8$-10$^9$ year timescale that stellar
and geothermal heat fluxes change. We will now examine this
assumption. For a planet with a net hydrogen source from the interior
(section~\ref{sec:source}), the adjustment timescale is
$\frac{4 a T_1^3(T_2-T_1)}{L_0 F}$. Let us assume that biological
consumption of H$_2$ is on the same scale as the interior flux
($L_0\approx P_0$) and set the hydrogen outgassing rate to a
reasonable value of $10^{10}$ H$_2$ molecules cm$^{-2}$ s$^{-1}$
\citep{haqq-misra2011}, or $3\times10^{-12}$~Pa~s$^{-1}$ for Earth's
gravity. Let us also assume that the solar plus geothermal heat flux
input ($F$) is on the order of 1~W~m$^{-2}$, $T_1$=300~K, and
$T_2-T_1$=100~K. With these assumptions, we arrive at a timescale of
$\approx10^6$~years, or a couple orders of magnitude smaller than the
timescale of changes in the forcings.

The feedbacks discussed in this letter clearly will not work for every
H$_2$-greenhouse planet. Among other issues, for some planets the
chemistry will not be correct, and for others there will be limited
amounts of necessary nutrients for biological functions. This is
analogous to the habitability of terrestrial planets in the Solar
System, only one of which appears to posses stabilizing feedbacks
capable of maintaining a habitable climate over long time periods.
Nevertheless, it is significant that stabilizing climate feedbacks on
an H$_2$-greenhouse planet result naturally merely from assuming the
existence of life processes that either consume or produce H$_2$ at a
rate that has a maximum as a function of temperature.

The key to detecting a stabilizing climate feedback on a planet with
an H$_2$-greenhouse planet would be to measure the H$_2$ surface
pressure (a difficult task) and compare it to what would be necessary
to maintain a habitable surface temperature. If a single planet were
found with the correct amount of H$_2$ and some sort of biosignature
\citep{Seager:2013fi}, this would be weak evidence that a stabilizing
climate feedback could exist on that planet. Alternatively, if it is
eventually found that an unexpectedly large number of H$_2$-greenhouse
planets orbiting far from their stars have the right amount of H$_2$
to support habitable surface conditions, a statistical argument could
be made that some sort of climate-stabilizing feedback (potentially
related to the ones described here) exists on those planets.

An interesting possibility is that an H$_2$-greenhouse planet might
actually exhibit both forms of climate stability discussed in this
letter over its lifetime. For example, a planet might receive a high
XUV flux from its star early in its life, causing significant H$_2$
escape to space, and maintain climate stability through H$_2$
production by life (section~\ref{sec:sink}). Later as the star becomes
less active, there might be a net source of H$_2$ to the atmosphere
from the interior, and the planet might maintain climate stability
through H$_2$ production by life (section~\ref{sec:source}).

I have neglected many processes in this letter. This is appropriate in
a letter that proposes a simple idea about a system for which there
are currently few constraints. Nevertheless, it might be interesting
to extend this work including consideration of the effects of clouds
and the effects of minor constituent gases on the atmospheric
temperature profile. Another issue that could be considered is how
atmospheric chemistry would affect the feedbacks described here
\citep[e.g.,][]{batalha2015testing}, although at distant orbits
photochemical effects would be minimal. Finally, more work could be
done biological issues such as the specific nutrients that might be
important to life on an H$_2$-greenhouse planet and how entropic waste
would be dealt with.

\clearpage

\section{Conclusions}
\label{sec:conclusions}

I have shown that stabilizing climate feedbacks on H$_2$-greenhouse
planets in distant orbits result from the simple assumption of the
existence of biological processes that either consume or produce H$_2$
at a rate that has a maximum as a function of temperature. Since this
assumption is not restrictive, it is possible that such planets could
remain habitable on geological timescales, making them more attractive
candidates for biosignature searches. I hope this work encourages more
thought on stabilizing feedbacks for the climate of distant planets
with habitable conditions supported by H$_2$ atmospheres.

\section{Acknowledgements}
I acknowledge support from an Alfred P. Sloan Research Fellowship and
from the NASA Astrobiology Institute’s Virtual Planetary Laboratory, which
is supported by NASA under cooperative agreement NNH05ZDA001C. I thank
Edwin Kite, Fred Ciesla, and Predrag Popovic for reading an early
draft of this letter and providing comments. I thank Jake Waldbauer
for helping me understand some basics of microbiology, Tim Cronin for
helping me think about the adjustment timescale, and Mohit Daswani for
explaining how serpentinization could effect the surface fluxes for
this problem. I thank the VPL team for giving me feedback on this work
in an internal seminar, especially Jim Kasting, Kevin Zahnle, and Norm
Sleep.

\end{document}